\documentclass[aps,prc,twocolumn,groupedaddress,showpacs]{revtex4}
\usepackage{graphicx}
\usepackage{amsmath}

\begin{document}


\title{
Phase diagram of nuclear ``pasta'' and its uncertainties
in supernova cores
}


\author{Hidetaka Sonoda$^{a,b}$, Gentaro Watanabe$^{c,d,b}$,
 Katsuhiko Sato$^{a,e,f}$, Kenji Yasuoka$^{g}$ and Toshikazu Ebisuzaki$^{b}$}
\affiliation{
$^{a}$Department of Physics, University of Tokyo, Tokyo 113-0033,
Japan
\\
$^{b}$The Institute of Chemical and Physical Research (RIKEN),
Saitama 351-0198, Japan
\\
$^{c}$CNR-INFM BEC Center, Department of Physics, University of Trento,
Via Sommarive 14, 38050 Povo (TN), Italy
\\
$^{d}$NORDITA, Blegdamsvej 17, DK-2100 Copenhagen \O, Denmark
\\
$^{e}$Research Center for the Early Universe, University of Tokyo,
Tokyo 113-0033, Japan\\
$^{f}$The Institute for the Physics and Mathematics of the Universe,
University of Tokyo,
Kashiwa, Chiba, 277-8568 Japan\\
$^{g}$Department of Mechanical Engineering, Keio University, Yokohama,
223-8522, Japan
}


\date{\today}

\begin{abstract}
We examine the model dependence of the phase diagram of inhomogeneous
nulcear matter in supernova cores using the quantum molecular
dynamics (QMD).
Inhomogeneous matter includes crystallized matter with nonspherical
nuclei -- ``pasta'' phases -- and the liquid-gas phase separating
nuclear matter.
Major differences between the phase diagrams 
of the QMD models can be 
explained by the energy of pure neutron matter at low densities
and the saturation density of asymmetric nuclear matter.
We show the density dependence of the symmetry energy is also useful
to understand uncertainties of the phase diagram.
We point out that, for typical nuclear models, 
the mass fraction of the pasta phases 
in the later stage of the collapsing cores is higher than
$10$--$20\, \%$.
\end{abstract}


\maketitle

\section{Introduction \label{intro}}

Terrestrial matter basically consists of spherical nuclei.
However, such an ordinary picture, that is, ``nuclei are spherical,''
might not be true in collapsing supernova cores just before bounce
and in the deepest region of inner crusts of neutron stars.
In dense matter close to the normal nuclear density
$\rho_0 \simeq 0.16 \, \mathrm{fm}^{-3}$,
nuclei would adopt nonspherical shapes including rod and slab.
Phases with these exotic nuclei are called ``pasta''
phases \cite{rpw,hashimoto}.

The pasta phases attract the attention of many researchers in the fields of
nuclear physics \cite{js_review}
and astrophysics \cite{bethe_rev,burrows_review,janka_review}.
Pasta phases have important astrophysical
effects on, e.g., neutrino opacity in supernova cores
\cite{horowitz,opacity}, neutrino emissivity in neutron star cooling
\cite{lrp,peth_thor,kaminker,gusakov}, etc.
Moreover, 
it has been shown in our previous work \cite{opacity} that the pasta phases
would occupy 10--20 \% of the mass of collapsing stellar core.
In such a case, the pasta phases could
have a remarkable impact on neutrino transport
in the core and hence success of supernova explosion.

Equilibrium states of the pasta phases have been investigated in many
earlier works (e.g., Refs.\ \cite{williams,lassaut,oyamatsu,
lrp,watanabe_liquid_nsm,watanabe_liquid,watanabe_liquid_err,maru_scr}).
These works have confirmed that,
with increasing density, nuclear shape basically
changes in the sequence sphere, rod, slab, rod-like bubbles,
spherical bubbles, and, finally, uniform nuclear matter 
(in some nuclear models, however, all of the above pasta phases do
not appear \cite{lrp,cheng,douchin,watanabe_liquid_nsm,watanabe_liquid,
watanabe_liquid_err}).
Although these earlier works have studied the phase diagram of the
pasta phases using various nuclear models,
the following two points are worth consideration.
First, in these works (except for Ref.\ \cite{williams}) authors
consider the above-mentioned specific nuclear structures and determine
the equilibrium state by comparing the free energy among them.
It is hardly possible to know in advance whether these assumed phases
include the true equilibrium state or there are other more stable states.
Thus, we have to examine how the phase diagram
is changed by relaxing this assumption.
Second, in collapsing supernova cores, 
where the pasta phases would appear, temperature reaches typically a
few MeV.  Thermal fluctuations on the nucleon distribution are not
completely negligible considering that the nucleon Fermi energy and
the nuclear binding energy are from several to tens MeV.  However,
thermal fluctuations cannot be properly incorporated by the framework
employed in the earlier works such as a liquid-drop model and the
Thomas-Fermi approximation.

To overcome the above problems, we use quantum
molecular dynamics (QMD) \cite{qmd_rc,qmd,qmd_hot,pasta_review}.
In these works, we have confirmed the pasta phases appear
at zero and nonzero temperatures and
the sequence of nuclear structures
with increasing density is the same as that in the earlier works.
There we have also obtained spongelike ``intermediate'' phases.
However, we have studied the pasta phases using 
only one specific nuclear force.  We note that uncertainties of nuclear force,
especially those of the surface energy and the symmetry energy,
have large effects on the phase diagram at subnuclear densities
\cite{watanabe_liquid_nsm,watanabe_liquid,watanabe_liquid_err,
oyamatsuiida}.
Thus, in the present work, we shall reveal the influence on the phase diagram
by uncertainties of nuclear force in the framework of QMD.

In the followings, we set the Boltzmann constant $k_B=1$.

\section{Framework of Quantum Molecular Dynamics
\label{QMD framework}}
\subsection{Models}
In our previous studies \cite{qmd_rc,qmd,qmd_hot,qmd_transition},
we used nuclear force developed by Maruyama {\it et al.} (Model 1)
\cite{maruyama} with medium equation-of-state (EOS) parameter set.
In the present work, we also use another model by Chikazumi {\it et al.}
(Model 2) \cite{chikazumi} to investigate the model dependence of phase
diagram.
The Hamiltonian of both the models is written as,
\begin{align}
\mathcal{H}=&K+V_{\mathrm{Pauli}}  \nonumber \\
 & +V_{\mathrm{Skyrme}}+V_{\mathrm{sym}}
   +V_{\mathrm{surface}}+V_{\mathrm{MD}}+V_{\mathrm{Coulomb}},
\end{align}
where $K$ is the kinetic energy; $V_{\mathrm{Pauli}}$ is the Pauli
potential, which is introduced to reproduce effects of
the Pauli exclusion principle;
$V_{\mathrm{Skyrme}}$ is the Skyrme-type interactions;
$V_{\mathrm{sym}}$ is the symmetry energy; $V_{\mathrm{surface}}$
is the potential dependent on the density gradient;
$V_{\mathrm{MD}}$ is the momentum-dependent potential in the form of
the exchange term of the Yukawa interaction;
and $V_{\mathrm{Coulomb}}$ is the Coulomb potential.
Each term is expressed as follows \cite{note_error}:
\begin{widetext}
\begin{eqnarray}
K&=&\sum_i \frac{{\bf P}_i^2}{2m_i},\\
V_{\mathrm{Pauli}}&=&\frac{1}{2}C_{\mathrm{P}} \left(
\frac{\hbar}{q_0p_0}\right)
\sum_{i,j(\neq i)} \exp \left[ -\frac{({\bf R}_i-{\bf R}_j)^2}{2q_0^2}
-\frac{({\bf P}_i-{\bf P}_j)^2}{2p_0^2}\right]\delta_{c_i c_j}
\delta_{\sigma_i \sigma_j},\\
V_{\mathrm{Skyrme}}&=&\frac{\alpha}{2\rho_0}\sum_{i,j(\neq i)}\rho_{ij}
+\frac{\beta}{(1+\tau)\rho_0^{\tau}}\sum_i \left[ \sum_{j(\neq i)}\int
d{\bf r}\, \tilde{\rho}_i({\bf r})\tilde{\rho}_j
({\bf r})\right]^{\tau},\\
V_{\mathrm{sym}}&=&\frac{C_\mathrm{s}}{2\rho_0}\sum_{i,j(\neq i)}
(1-2|c_i-c_j|)\rho_{ij},\\
V_{\mathrm{surface}}&=&\frac{V_{\mathrm{SF}}}{2\rho_0^{5/3}}
\sum_{i,j(\neq i)}\int d{\bf r}\,
\nabla \rho_i({\bf r})\cdot \nabla \rho_j({\bf r}),\\
V_{\mathrm{MD}}&=&\frac{C_{\mathrm{ex}}^{(1)}}{2\rho_0}\sum_{i,j (\neq
 i)} \frac{1}{1+\left[\frac{{\bf P}_i-{\bf P}_j}{\hbar\mu_1}\right]^2}
\, \rho_{ij}+
\frac{C_{\mathrm{ex}}^{(2)}}{2\rho_0}\sum_{i,j (\neq
 i)} \frac{1}{1+\left[\frac{{\bf P}_i-{\bf P}_j}{\hbar\mu_2}\right]^2}
\, \rho_{ij},\\
V_{\mathrm{Coulomb}}&=&\frac{e^2}{2}\sum_{i,j(\neq i)}c_i
c_j \int\int
d{\bf r}d{\bf r}'\frac{1}{|{\bf r}-{\bf r}'|}\rho_i({\bf r})
\rho_j({\bf r}'),\label{eq_exp_int}
\end{eqnarray}
\end{widetext}
where ${\bf R}_i$ and ${\bf P}_i$ are the centers of position and momentum of 
the wave packet of $i$th nucleon and
$m_i$, $\sigma_i$, and $c_i$ ($c_i$=1 for protons and $c_i$=0
for neutrons)
denote the mass, the spin, and the
electric charge (in units of $e$) of $i$th nucleon.
Here $\rho_{ij}$ means the overlap between the densities of
$i$th and $j$th nucleons,
\begin{eqnarray}
\rho_{ij}=\int d{\bf r}\, \rho_i({\bf r})\rho_j({\bf r}),
\end{eqnarray}
and the single-nucleon densities $\rho_i({\bf r})$ and
$\tilde{\rho}_i({\bf r})$ are given by
\begin{eqnarray}
\rho_i({\bf r})=\frac{1}{(2\pi L_\mathrm{w}^2)^{3/2}} \exp \left[
 -\frac{({\bf r}-{\bf R}_i)^2}{2L_\mathrm{w}^2}\right],\\
\tilde{\rho}_i({\bf r})=\frac{1}{(2\pi \tilde{L}_\mathrm{w}^2)^{3/2}}
 \exp \left[ -\frac {({\bf r}-{\bf R}_i)^2}
 {2\tilde{L}_\mathrm{w}^2}\right],
\end{eqnarray}
with the normal width $L_{\mathrm{w}}$ and
the modified width $\tilde{L}_\mathrm{w}$ of the wave packet,
\begin{eqnarray}
 \tilde{L}_\mathrm{w}^2=\frac{(1+\tau)^{1/\tau}}{2}
  L_\mathrm{w}^2 \label{def_l_tilde}
\end{eqnarray}
(The squared widths $L_{\rm w}^2$ and $\tilde{L}_\mathrm{w}^2$
correspond to $L$ and $\tilde{L}$, respectively, in the notation of 
Refs.\ \cite{maruyama,chikazumi}.)

Parameters for the models are shown in Table \ref{para_set}.
Note that $V_{\mathrm{SF}}=0$ for model 1, i.e., model 1 does not
include $V_{\mathrm{surface}}$.
Model parameters $q_0$, $p_0$, and $C_\mathrm{P}$
in the Pauli potential are determined by fitting
the kinetic energy of free Fermi gas at zero temperature.
The other model parameters are determined to reproduce
the saturation properties of symmetric nuclear matter [i.e., 
the saturation density ($\simeq 0.16\, \mathrm{fm}^{-3}$),
saturation energy ($-16$ MeV per baryon), and incompressibility
(280 MeV)], and the binding energy and rms radius of
the ground state of stable nuclei.
Especially these properties of heavy nuclei are better reproduced by
model 2 than by model 1 due to the term $V_{\mathrm{surface}}$
\cite{kido}.
Note that $V_{\mathrm{surface}}$ is just a potential depending on
the density gradient and is different from the surface energy.
The surface energy comes from an energy loss
due to the deficiency of nucleons interacting with each other 
in the region of the nuclear surface.

\begin{table}
\caption{Parameter sets for model 1 \cite{maruyama} and model 2
 \cite{chikazumi}.
}
\label{para_set}
\begin{tabular}{c|cc}
& $\qquad$Model 1$\qquad$ &$\qquad$ Model 2 $\qquad$\\ \hline
$C_{\mathrm{P}}$ (MeV) & 207 & 115 \\
$p_0$ (MeV/$c$) & 120 & 120 \\
$q_0$ (fm) & 1.644 & 2.5 \\
$\alpha$ (MeV) & $-92.86$ & $-121.9$ \\
$\beta$ (MeV) & 169.28 & 197.3 \\
$\tau$ & 1.33333 & 1.33333 \\
$C_{\mathrm{s}}$ (MeV) & 25.0 & 25.0 \\
$V_{\mathrm{SF}}$ (MeV) & 0 & 20.68 \\
$C_{\mathrm{ex}}^{(1)}$ (MeV) & $-258.54$ & $-258.54$ \\
$C_{\mathrm{ex}}^{(2)}$ (MeV) & 375.6 & 375.6 \\
$\mu_1$ ($\mathrm{fm}^{-1}$) & 2.35 & 2.35 \\
$\mu_2$ ($\mathrm{fm}^{-1}$) & 0.4 & 0.4 \\
$L_{\mathrm{w}}^2$ ($\mathrm{fm}^2$) & 2.1 & 1.95 \\
$\rho_0$ ($\mathrm{fm}^{-3}$) & 0.165 & 0.168 \\ \hline
\end{tabular}
\end{table}

\subsection{Equations of motion \label{sec:eom}}
We show equations of motion of QMD, which we employ to simulate 
the equilibrium states at zero and nonzero temperatures.
The Hamiltonian form of the QMD equations of motion is written as
\begin{eqnarray}
 \begin{array}{ccr}
 \dot{{\bf R}}_i&=&\displaystyle{\frac{\partial \mathcal{H}}{\partial
 {\bf P}_i}}, \\
  & & \\
 \dot{{\bf P}}_i&=&-\displaystyle{\frac{\partial \mathcal{H}}{\partial
 {\bf R}_i}}.
 \end{array}\label{QMDeq}
\end{eqnarray}
We cool down the system using the following equations of motion 
in which we introduce extra friction terms to the above equations
\cite{maruyama}:
\begin{eqnarray}
 \begin{array}{ccr}
 \dot{{\bf R}}_i&=&\displaystyle{\frac{\partial \mathcal{H}}{\partial
 {\bf P}_i}}
 -\xi_R \frac{\partial \mathcal{H}}{\partial {\bf R}_i},\\
 & & \\
 \dot{{\bf P}}_i&=&-\displaystyle{\frac{\partial \mathcal{H}}{\partial
 {\bf R}_i}}
 -\xi_P \frac{\partial \mathcal{H}}{\partial {\bf P}_i}.
 \end{array}\label{friceq}
\end{eqnarray}
Here, the friction coefficients $\xi_R$ and $\xi_P$ are positive
definite, which determine the relaxation time scale and 
lead to a monotonic decrease of the total energy.

Instead of the normal kinetic temperature, which loses
its physical meaning for the system with momentum-dependent potentials,
we use effective temperature $T_{\mathrm{eff}}$ 
proposed by Ref.\ \cite{chikazumi}:
\begin{eqnarray}
 \frac{3}{2} T_{\mathrm{eff}}=\frac{1}{\mathcal{N}}\sum_{i} \frac{1}{2}
 {\bf P}_i \cdot \frac{d{\bf R}_i}{dt},\label{defTeff}
\end{eqnarray}
where $\mathcal{N}$ is the total number of particles.
If we perform Metropolis Monte Carlo simulations
with the setting temperature $T_{\mathrm{set}}$,
the long-time average of the effective temperature coincides with
$T_{\mathrm{set}}$ quite well \cite{qmd_hot}.
This shows $T_{\mathrm{eff}}$ is consistent with
the temperature in the Boltzmann statistics.

To obtain the equilibrium state at finite temperatures, 
we use the Nos\'{e}-Hoover thermostat \cite{nose1,nose2,hoover}
modified for momentum-dependent potentials
\cite{qmd_hot,nosehoover_perez}.
The Hamiltonian of the system with this thermostat is
\cite{erratum}
\begin{eqnarray}
 \mathcal{H}_{\mathrm{Nose}}=\sum_{i} \frac{{\bf P}_i^2}{2m_i}
 +\mathcal{U}(\{{\bf R}_i,{\bf P}_i\})+\frac{s^2p_s^2}{2Q}
 +g\frac{\ln s}{\beta}.\label{nosehamiltonian}
\end{eqnarray}
Here $\mathcal{U}$ is the momentum-dependent potential, $s$ is the additional
dynamical variable for time scaling, $p_s$ is the momentum conjugate
to $s$, $Q$ is the thermal inertial parameter corresponding to
a coupling constant between the system and the thermostat,
$g$ is a parameter to be determined as $3N$ by the condition for
generating the canonical ensemble in the classical molecular dynamic
simulations, and $\beta$ is the reciprocal of
$T_{\mathrm{set}}$ of the thermostat.
Then equations of motion are
\begin{eqnarray}
 \begin{array}{ccl}
  \displaystyle{\frac{d{\bf R}_i}{dt}}&=&
   \displaystyle{\frac{{\bf P}_i}{m_i}}
   +\displaystyle{\frac{\partial \mathcal{U}}{\partial {\bf P}_i}},\\
  & & \\
  \displaystyle{\frac{d{\bf P}_i}{dt}}&=&-
   \displaystyle{\frac{\partial \mathcal{U}}
   {\partial {\bf R}_i}}-\xi {\bf P}_i,\\
  & & \\
  \displaystyle{\frac{1}{s}\frac{ds}{dt}}&=&\xi,\\
  & & \\
  \displaystyle{\frac{d\xi}{dt}}&=&
   \displaystyle{\frac{1}{Q}}\left\{\sum_{i}
   \left( \displaystyle{\frac{{\bf P}_i^2}{m_i}}
   +{\bf P}_i\cdot\displaystyle{\frac{\partial\mathcal{U}}{\partial {\bf
   P}_i}} \right)-\displaystyle{\frac{g}{\beta}} \right\},
 \end{array}
\end{eqnarray}
with
\begin{eqnarray}
 \xi\equiv \frac{sp_s}{Q},
\end{eqnarray}
where $\xi$ means the thermodynamic friction coefficient.
During the time evolution described by the above equations, 
$\mathcal{H}_\mathrm{Nose}$ is
conserved and $T_{\mathrm{eff}}$ fluctuates around $T_{\mathrm{set}}$.

\section{Pure neutron matter and stable nuclei \label{sec:pnm_nuc}}
To understand the properties of our QMD models,
we first calculate the energy and the proton chemical potential of pure neutron
matter at zero temperature.
In addition, we investigate the surface diffuseness and the 
surface tension of stable nuclei.
These are one of the key uncertainties that affects the phase diagram at 
subnuclear densities \cite{watanabe_liquid_nsm,
watanabe_liquid,watanabe_liquid_err,oyamatsuiida}.
To obtain pure neutron matter at zero temperature, 
we use the frictional relaxation method [Eq.\ (\ref{friceq})]
with the cooling time scale of $\mathcal{O}(10^3)\, \mathrm{fm}/c$.
We calculate the proton chemical potential at zero temperature 
from the change of the energy 
by inserting a proton into the pure neutron matter.
Here we relax the position and momentum of the proton with fixing those 
of neutrons (for more details about the procedures, see Ref.\ \cite{qmd}).
To obtain the ground state of finite nuclei, we use the conjugate gradient
method \cite{recipe}.

\begin{center}
\begin{figure*}
\rotatebox{270}{
\resizebox{!}{12cm}
{\includegraphics{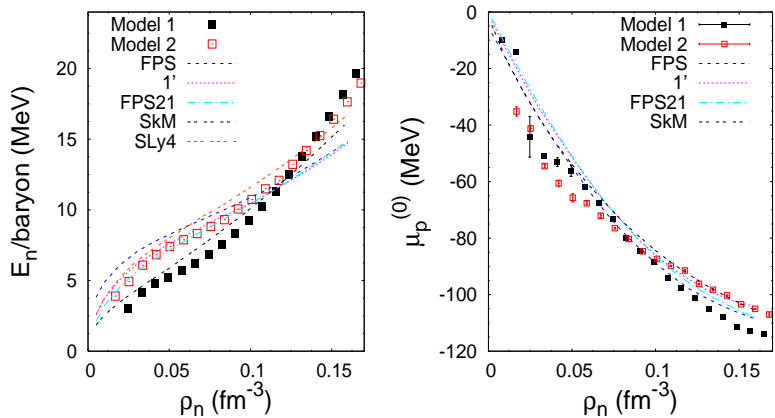}}}
\caption{(Color online) Energy per baryon (left panel) and proton
 chemical potential
 $\mu_p^{(0)}$ (right panel) in pure neutron matter
 calculated by the QMD models and several other nuclear forces.
 The line denoted by SLy4 is from Ref.\
 \cite{douchin}. The other lines (FPS, 1', FPS21, and SkM) are
 from Ref.\ \cite{nucl_mod_prl}. \label{pro_neu_mat}}
\end{figure*}
\end{center}

Energy $E_{n}$ per baryon of pure neutron matter is shown
in the left panel of Fig.\ \ref{pro_neu_mat}.
At subnuclear densities, $E_n$ of both the QMD models exhibits
reasonable values compared with those of other nuclear models.
At lower densities of $\rho_{n}\alt 0.1$ fm$^{-3}$, 
model 1 gives relatively small energy but close to SkM,
which gives the lowest energy among the other models.
At densities below $0.12 \, \mathrm{fm}^{-3}$, the energy of pure neutron
matter for model 2 is larger than that for model 1.
This tends to prevent neutrons from dripping out of nuclei.
As we will see later, the number density of dripped neutrons for model 2
is indeed smaller than that for model 1.

According to Ref.\ \cite{oyamatsuiida}, 
parameter $L$ of the density-dependent symmetry energy coefficient
also plays an important role
in determining the density region of the pasta phases.
This parameter is directly related to the derivative of
the energy of pure neutron matter with respect to $\rho_n$ 
at the normal nuclear density [see Eq. (4) in Ref.\ \cite{oyamatsuiida}]:
\begin{equation}
  L=3\rho_0\frac{\partial}{\partial\rho_n}\left(\frac{\epsilon_n}{\rho_n}\right)_{\rho_0},
\label{L}
\end{equation}
where $\epsilon_n$ is the energy density of pure neutron matter.
The left panel of Fig.\ \ref{pro_neu_mat} shows a larger slope
of the energy for model 1 than model 2, which leads to a larger value
of $L$ for model 1.
From Eq.\ (\ref{L}), we obtain $L=93$ MeV for model 1 
and $L=80$ MeV for model 2.
This difference affects
the density at which matter becomes uniform as we will discuss 
in the next section.

In the right panel of Fig.\ \ref{pro_neu_mat}, 
we show the proton
chemical potential $\mu_p^{(0)}$ in pure neutron matter at zero temperature 
calculated by the QMD models together with those by other nuclear models.
This result shows QMD model 1 gives slightly lower values of
$\mu_p^{(0)}$ at high densities compared with other nuclear models,
whereas model 2 gives lower values at low densities.
As discussed in Refs.\ \cite{watanabe_liquid,watanabe_liquid_err},
uncertainty of $\mu_p^{(0)}$
little affects the phase diagram of supernova matter 
because the number density of dripped neutrons
is very small \cite{note_chempot}.

\begin{center}
\begin{table*}
\caption{Several quantities of typical heavy nuclei calculated by
 each QMD model. $E_{B}/A$ is the empirical binding energy
 from Ref.\ \cite{nucl_emp_data}.
 $E/A$ is the binding energy calculated by QMD.
 $\rho_{\mathrm{in}}$ is the central nucleon density of the nucleus.
 $b_p$ and $b_n$ are the surface diffuseness parameter for protons and
 neutrons, respectively, defined as Eq.\ (\ref{def_b}).
 $W$ is the binding energy defined as Eq.\ (\ref{def_w}) evaluated for
 the central density and a proton fraction of the nucleus.
 $\sigma$ is the surface tension defined as Eq.\ (\ref{def_sigma}).
 \label{tab:nuclei}}
\begin{tabular}{|c||c|c|c|c|c|c|c|c|}
\hline
 & \multicolumn{2}{c|}{$^{56}\mathrm{Fe}$} &
 \multicolumn{2}{c|}{$^{90}\mathrm{Zr}$} &
 \multicolumn{2}{c|}{$^{208}_{82}\mathrm{Pb}$}&
 \multicolumn{2}{c|}{$^{238}_{92}\mathrm{U}$}\\ \hline \hline
$E_{B}/A$ (MeV)&
 \multicolumn{2}{c|}{$-8.79$} &
 \multicolumn{2}{c|}{$-8.71$} &
 \multicolumn{2}{c|}{$-7.87$} &
 \multicolumn{2}{c|}{$-7.57$}  \\ \hline \hline
Model & 1 & 2 & 1 & 2 & 1 & 2 & 1 & 2 \\ \hline
$E/A$ (MeV)& $-9.09$ & $-8.77$ & $-9.25$ & $-8.68$ & $-8.66$ &
 $-7.65$ & $-8.31$ & $-7.38$ \\ \hline
$\rho_{\mathrm{in}}$ ($\mathrm{fm}^{-3}$)& 0.226& 0.215& 0.213&
 0.211& 0.193 & 0.163 & 0.173 & 0.160\\ \hline
$b_p$ (fm)& 4.0 & 3.2 & 4.2 & 3.2 & 3.4 & 3.1 & 3.3 & 3.0 \\ \hline
$b_n$ (fm)& 4.1 & 3.3 & 4.0 & 3.5 & 3.8 & 3.4 & 3.8 & 3.4 \\ \hline
$W(\rho_{\mathrm{in}},x_p)$ (MeV)& $-14.11$ & $-14.80$ & $-14.47$ &
 $-14.72$ & $-14.05$ & $-14.51$ & $-14.18$ & $-14.26$ \\ \hline
$\sigma$ (MeV $\mathrm{fm}^{-2}$)& 0.66 & 1.13 & 1.03 & 1.09 &
 0.62 & 1.07 & 0.73 & 1.03 \\ \hline
\end{tabular}
\end{table*}
\end{center}

In Table \ref{tab:nuclei}
we show several quantities related to the surface diffuseness
and the surface energy of typical heavy nuclei, 
$^{56}\mathrm{Fe}$, $^{90}\mathrm{Zr}$, $^{208}\mathrm{Pb}$,
and $^{238}\mathrm{U}$,
calculated for each QMD model.
We calculate the surface diffuseness parameter, which,
following the spirit of Ref.\ \cite{nemeth}, we define as
\begin{eqnarray}
b_{i}\equiv \frac{\rho_{i,\mathrm{in}}}
 {|d\rho_{i}/dr|_{\mathrm{max}}}\quad (i=p,n). \label{def_b}
\end{eqnarray}
Here we have replaced $\rho_0$ in the definition of Ref.\
\cite{nemeth}, which is employed for the semi-infinite system,
by the central density $\rho_{i,{\rm in}}$ of the finite nucleus.

We estimate the surface energy $\sigma$ within the framework of QMD 
by subtracting the contributions of the bulk and the Coulomb energies
from the total binding energy $E$:
\begin{eqnarray}
\sigma =\frac{E-E_{\mathrm{coul}}
-AW(\rho_{\mathrm{in}},x_p)}{4\pi R^2}.
\label{def_sigma}
\end{eqnarray}
Here $A$ and $x_p$ are the mass number and the proton fraction
of the nucleus, $E_{\mathrm{coul}}$ is the Coulomb energy of the
nucleus, and $W(\rho_{\mathrm{in}},x_p)$ is the bulk
energy evaluated for the central density $\rho_{\rm in}$ of the nucleus.
For $W(\rho_{\mathrm{in}},x_p)$, we assume the following form, 
\begin{eqnarray}
W(\rho_{\mathrm{in}},x_p)\equiv & W_V
+\displaystyle{\frac{1}{2}}K_0\left( 1-
\displaystyle{\frac{k_{\mathrm{in}}}{k_0}}\right)^2 \nonumber \\
&+4S_V \left( x_p-\displaystyle{\frac{1}{2}}\right)^2, \label{def_w}
\end{eqnarray}
where $W_V$ is the binding energy of symmetric nuclear matter at
$\rho_0$,
$K_0$ is the incompressibility, and
$k_{\mathrm{in}}\equiv (3\pi^2\rho_{\rm in}/2)^{1/3}$ and
$k_{0}\equiv (3\pi^2\rho_0/2)^{1/3}$ 
are the wave numbers of nucleon at $\rho=\rho_{\mathrm{in}}$
and $\rho_0$, respectively.
We set $W_V=-16$ MeV, $K_0=280$ MeV, $S_V=34.6$ MeV, and
$\rho_0=0.165 \, \mathrm{fm}^{-3}$ for model 1 \cite{maruyama}
and  $W_V=-16$ MeV, $K_0=280$ MeV, $S_V=33$ MeV, and
$\rho_0=0.168 \, \mathrm{fm}^{-3}$ for model 2 \cite{kido,chikazumi}.
The nuclear radius $R$ in Eq.\ (\ref{def_sigma}) is defined as
\begin{eqnarray}
R\equiv \left( \frac{3}{4\pi} \frac{A}{\rho_{\mathrm{in}}}\right)^{1/3}.
\end{eqnarray}
Table \ref{tab:nuclei} shows that the
surface energy $\sigma$ estimated for model 2 is systematically higher
than that for model 1.
We also see that the surface diffuseness parameters $b_i$
of both neutrons and protons are smaller for model 2 than model 1,
which means model 2 yields steeper density profile of the
nuclear surface.
Both of these two facts consistently indicate that 
the nuclear surface energy for model 2 is greater than that for model 1.

\section{Nuclear matter at subnuclear densities\label{sec:result}}
\subsection{Simulations and snapshots \label{sec:proc}}
We studied the $(n,p,e)$ system of the proton fraction $x_p=0.3$
at zero and nonzero temperatures.
The value of $x_p=0.3$ is typical for matter in collapsing supernova cores.
For this purpose, we use 2048 nucleons (614 protons and 1434 neutrons) 
in a cubic simulation box with periodic boundary condition
\cite{note_number}.
We assume that the system is not magnetically polarized, i.e.,
it contains equal numbers of protons (and neutrons) with
spin up and spin down.
The relativistic degenerate electrons can be well approximated as
a uniform background in our situations \cite{bbp,pr_review,screening}.
The Coulomb interaction is calculated by the Ewald sum 
(expressions used in our simulations are given in Ref.\ \cite{qmd}).
To obtain equilibrium states
both at zero and nonzero temperatures,
we perform simulations in the following procedure.
We first prepare nuclear matter at $T=10$ MeV by equilibrating the
system for about $3000 \, \mathrm{fm}/c$
using the Nos\'{e}-Hoover thermostat.
To reproduce the ground state, we cool down the system with the frictional
relaxation (\ref{friceq}) 
for the time scale of $\mathcal{O}(10^{3-4}) \, \mathrm{fm}/c$.
This time scale is sufficiently larger than the relaxation time scale
of the system $\mathcal{O}(10^2)$ fm/$c$,
which is determined by the typical length scale of
the structure $\mathcal{O}(10)$ fm and the sound velocity 
$\mathcal{O}(10^{-1})\, c$.
To obtain nuclear matter at some fixed nonzero temperature of 
$T_{\mathrm{set}}$, we start from a snapshot with the effective temperature 
$T_{\mathrm{eff}} \simeq T_{\mathrm{set}}$ obtained in the
above cooling process.
We then equilibrate it with the Nos\'{e}-Hoover thermostat for at least
$5000 \, \mathrm{fm}/c$.

\begin{center}
\begin{figure*}
\rotatebox{270}{
\resizebox{!}{15cm}
{\includegraphics{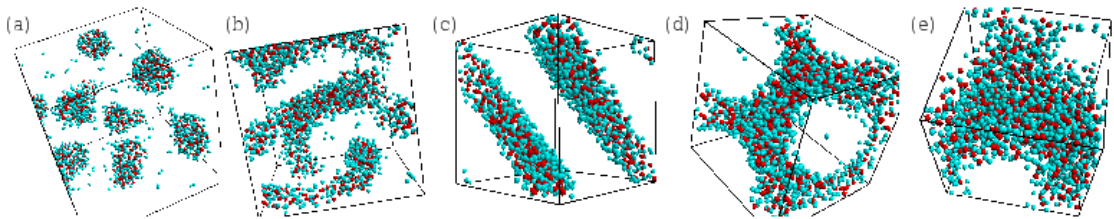}}}
\caption{(Color online)
 Nucleon distribution of the pasta phases at zero temperature
 for QMD model 2.
 Simulations are performed with 2048 nucleons at a
 proton fraction $x_p=0.3$.
 Each red (blue) particle corresponds to a proton (neutron).
 Each picture shows the pasta phase with
 (a) spherical nuclei ($0.1\rho_0=0.0168 \, \mathrm{fm}^{-3}$),
 (b) cylindrical nuclei ($0.2\rho_0=0.0336 \, \mathrm{fm}^{-3}$),
 (c) slablike nuclei ($0.393\rho_0=0.0660 \, \mathrm{fm}^{-3}$),
 (d) cylindrical holes ($0.49\rho_0=0.0823 \, \mathrm{fm}^{-3}$), and 
 (e) spherical holes ($0.575\rho_0=0.0966 \, \mathrm{fm}^{-3}$).
 Box sizes are (a) 49.58 fm, (b) 39.35 fm, (c) 31.42 fm, (d) 29.19 fm,
 and (e) 27.67 fm.
\label{fig_cold_pic}}
\end{figure*}
\end{center}

In Fig.\ \ref{fig_cold_pic}, we show nucleon distributions of the pasta
phases at zero temperature obtained by the simulations for model 2.
Compared with those for model 1 (see, e.g., Fig.\ 2 in
Ref.\ \cite{qmd_rc}), less dripped neutrons are observed.
Especially, dripped neutrons almost disappear when nuclei become
planar.
This would be due to the higher energy $E_n$ of pure neutron matter
at low densities as shown in Fig.\ \ref{pro_neu_mat}.
As stated later in Sec.\ \ref{sec:phdig},
this relative lack of dripped neutrons
decreases the density at which the fission instability of 
spherical nuclei occurs.

\begin{center}
\begin{figure*}
\rotatebox{270}{
\resizebox{!}{15cm}
{\includegraphics{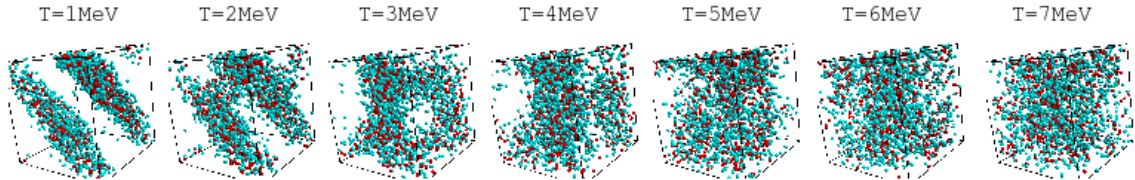}}}
\caption{(Color online) Nucleon distribution at $T=1$--7 MeV 
and a fixed density for model 2.
Simulations are performed with 2048 nucleons at 
$x_p=0.3$ and $\rho=0.393\rho_0=0.0660 \, \mathrm{fm}^{-3}$
(the box size is 31.42 fm), where
the phase with slablike nuclei is obtained at zero temperature.
Each red (blue) particle corresponds to a proton (neutron).
\label{fig_hot_pic}}
\end{figure*}
\end{center}

In Fig.\ \ref{fig_hot_pic} we show how slablike
nuclei melt to be uniform nuclear matter with increasing
temperature.
We can see a basic picture of phase transitions at nonzero temperatures
for fixed densities from this figure.
Suppose we increase the temperature from zero.
At $T\simeq 1$ MeV, neutrons start to evaporate from nuclei.
At $T=2$--3 MeV, increase of the volume fraction of nuclear matter 
region due to its thermal expansion
triggers a transition of the nuclear structure: slablike nuclei
start to connect with each other to form cylindrical bubbles.
At $T\simeq 4$ MeV, protons cannot be completely confined in nuclei and
start to evaporate.
In this situation, nuclear surface can no longer be identified,
but in many cases there remains clustering of protons and neutrons,
i.e., nuclear matter does not completely become uniform.
Finally, the clustering inside nuclear matter completely disappears
and matter becomes uniform.
We observe this transition at $T=6$--7 MeV 
in the case of Fig.\ \ref{fig_hot_pic}.

\subsection{Identification of phases \label{sec:def_ph}}
As can be seen from the snapshots of the nucleon distribution in the
previous section,
especially at nonzero temperatures, nuclei have complicated shapes
and moreover nuclear surface becomes diffuse.
We thus need to quantitatively identify the nuclear shape
from nucleon distributions obtained by the simulations.
For this purpose, we use the two-point correlation
function and the Minkowski functionals, especially the area-averaged
integral mean curvature $\langle H \rangle $
and the Euler characteristic density $\chi /V$.
The Euler characteristic $\chi$ is a purely topological quantity
and is expressed as
\begin{align}
\chi =& (\mathrm{number \, of \, isolated \, regions})-
(\mathrm{number \, of \, tunnels}) \nonumber\\
&+(\mathrm{number \, of \, cavities}).
\label{expl_chi}
\end{align}
For detailed procedures of calculating the Minkowski functionals,
see Ref.\ \cite{qmd} (see also Refs.\ \cite{michielsen,gott,weinberg}
for the algorithm of the calculation).

With these quantities, we can completely classify
the following typical pasta phases:
\begin{eqnarray}
 \left\{
  \begin{array}{cl}
    \langle H \rangle >0 \; \mathrm{and} \; \chi /V >0&
     \mathrm{for}\; \mathrm{spherical \, nuclei \, (SP)}\\
    \langle H \rangle >0 \; \mathrm{and} \; \chi /V =0&
     \mathrm{for}\; \mathrm{cylindrical \, nuclei \, (C)}\\
    \langle H \rangle =0 \; \mathrm{and} \; \chi /V =0&
     \mathrm{for}\; \mathrm{slablike \, nuclei \, (S)}\\
    \langle H \rangle <0 \; \mathrm{and} \; \chi /V =0&
     \mathrm{for}\; \mathrm{cylindrical \, holes \, (CH)}\\
    \langle H \rangle <0 \; \mathrm{and} \; \chi /V >0&
     \mathrm{for}\; \mathrm{spherical \, holes \, (SH)}.
  \end{array}
 \right. \label{def_ph}
\end{eqnarray}
As shown above, $\chi /V$ is always positive or zero for these phases.
However, in our previous studies \cite{qmd,qmd_hot}, and also in the
present study, we obtain ``spongelike'' phases with multiply
connected structures characterized by $\chi /V <0$.
We call the phases with $\chi /V <0$
intermediate phases, which appear
in the density region between those of the phases with rodlike
nuclei and slablike nuclei [here we denote as (C,S)],
and between slablike nuclei and rod-like bubbles
[denote as (S,CH)].
The former phase (C,S) gives $\langle H \rangle >0$ and
$\chi /V <0$ and the latter one, (S,CH), gives $\langle H \rangle <0$
and $\chi /V<0$.
Considering similarity of exotic structures observed in nuclear
matter and diblock copolymers,
several authors pointed out a possibility of more complex structures
than ordinary pasta structures, i.e., rods or slabs
\cite{pr_review,pasta_review,ohta_review}.
In diblock copolymer melts, one experimentally finds
complicated structures, e.g., so-called gyroid (G) structure 
and the ordered bicontinuous double diamond (OBDD) structure.
Although the intermediate phases obtained in our studies are different
from G and OBDD phases, it is possible that some complicated
structure other than one-dimensional lattice of slablike nuclei,
hexagonal lattice of rodlike ones, and BCC lattice of spherical
ones appears \cite{pr_review,pasta_review}.

The quantities $\langle H \rangle$ and $\chi /V$
can be calculated only if nuclear surface
can be identified. 
Suppose we increase the temperature, nuclei start to melt
and nuclear surface cannot be necessarily identified.
In this situation, the density inhomogeneity of long wavelength
starts to be smoothed out but still remains.
Further increasing temperature and exceeding some critical point,
matter becomes uniform.
To determine the boundary where inhomogeneity disappears,
we use the two-point correlation function defined as
\begin{eqnarray}
\xi_{ii} (r) &=& \frac{1}{4\pi}\int d\Omega_{\bf r} \, \frac{1}{V}
\int d{\bf x} \, \delta_i ({\bf x})\delta_i ({\bf x}+{\bf r})\\
&\equiv& \langle \delta_i ({\bf x})\delta_i ({\bf x+r})
\rangle_{{\bf x},\Omega_{\bf r}}. \label{def_xiii}
\end{eqnarray}
Here $i$ specifies the species of particles 
(proton or neutron, or both collectively),
$\langle \cdots \rangle_{{\bf x},\Omega_{\bf r}}$ denotes
an average over the position ${\bf x}$ and the direction of ${\bf r}$,
and $\delta_i({\bf x})$ is the fluctuation of the density field
$\rho^{(i)}({\bf x})$ given by
\begin{eqnarray}
\delta_i({\bf x})\equiv \frac{\rho^{(i)}({\bf x})-\overline{\rho^{(i)}}}
{\overline{\rho^{(i)}}}, \label{def_delta}
\end{eqnarray}
with the average density of protons or neutrons or both collectively:
\begin{eqnarray}
\overline{\rho^{(i)}}\equiv \frac{\mathcal{N}_i}{V}.\label{def_rhobar}
\end{eqnarray}
If the system of nuclear matter is separated into liquid and gas phases,
the two-point correlation function $\xi_{NN}(r)$ of nucleons
oscillates around zero even at long distances.
Otherwise, $\xi_{NN}(r)$ is almost zero (for $r\gtrsim 7$ fm
in our cases) without osillating.
In this case we judge the system is uniform (see Ref.\ \cite{qmd_hot}
for more details).

\subsection{Phase diagrams\label{sec:phdig}}

\begin{figure}
\rotatebox{270}{
\resizebox{6cm}{!}
{\includegraphics{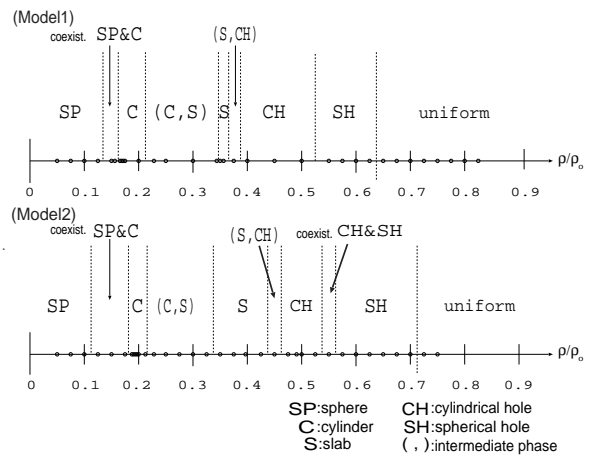}}}
\caption{
 Phase diagram of the pasta phases
 at zero temperature for model 1 (upper panel) and model 2 (lower
 panel). Proton fraction is $x_p=0.3$.
 Horizontal axis is normalized in units of the normal nuclear density
 $\rho_0$ for each model. 
 Abbreviations SP, C, S, CH, and SH mean phases with spherical nuclei,
 cylindrical nuclei, slablike nuclei, cylindrical holes,
 and spherical holes, respectively.
 The parentheses (A,B) show an intermediate phase between A and B phases.
\label{phdig_cold}}
\end{figure}

In Fig.\ \ref{phdig_cold} we show the phase diagram of the pasta phases
at zero temperature calculated by QMD.
The upper and the lower panels of Fig.\ \ref{phdig_cold} are
the phase diagram for model 1 and 2, respectively.
The sequence of nuclear shapes with increasing density
is the same as that predicted by
all the previous works including those by QMD.
The density region of the phases with nonspherical nuclei
for model 2 is larger than that for model 1:
spherical nuclei begin to elongate at a lower density
and spherical bubbles remain until a higher density for model 2.

The decrease of the density at which nuclei start to be deformed
would be due to the smaller number density of dripped neutrons
for model 2 compared with model 1, which is originated from
the larger energy of pure neutron matter at low densities shown
in Fig.\ \ref{pro_neu_mat}.
Nuclei are more neutron rich due to the smaller
number density of dripped neutrons.
The nucleon number density inside the nuclei is relatively small
because of the smaller saturation density of nuclear matter
at the lower proton fraction.
Thus the volume fraction of the nuclei increases, which decreases
the fission instability of spherical nuclei \cite{pr_review}.
Furthermore, the decrease of the number density of dripped neutrons
increases the mass number of nuclei, which also increases
their volume fraction.
As a result, although the saturation density of asymmetric nuclear
matter of $x_p=0.3$ for model 2 is higher (see below),
the volume fraction of nuclei for model 2
is about 10 \% higher than that for model 1
at $\rho=0.1 \, \rho_0$.
Thus the density at which the fission instability occurs for model 2
should be $\sim 0.01 \rho_0$ lower than that for model 1.
This value is consistent with our present results,
which indicate the difference between the two models
is smaller than $0.05\rho_0$.

As to the boundary between the regions of the phase with spherical holes 
and of the uniform phase, however, the increase of the density
$\rho_{\rm m}$ at which matter becomes uniform 
would be due to higher saturation density
at a proton fraction $x_p=0.3$ for model 2.
The saturation density and the saturation energy per baryon
at $x_p=0.3$ are
$0.136 \, \mathrm{fm}^{-3}$ and $-12.01$ MeV for model 1,
and
$0.147 \, \mathrm{fm}^{-3}$ and $-9.86$ MeV
for model 2 \cite{note_saturation}.
The saturation density directly affects $\rho_{\rm m}$ 
and, consequently, the $\rho_{\rm m}$ of model 2 is higher than that
of model 1.
Here one should keep in mind an effect of the surface energy.
In Sec.\ \ref{sec:pnm_nuc}, we show that model 2 gives a larger
surface tension compared with model 1. 
The larger surface tension of model 2 favors the uniform phase
without bubbles and acts to decrease $\rho_{\rm m}$. 
However, this effect should be small compared to the contribution of 
the saturation density, which can be understood by taking account of 
the incompressible property of nuclear matter; in the incompressible limit,
$\rho_{\rm m}=\rho_{\rm s}$ and the surface tension does not affect 
$\rho_{\rm m}$ at all.

The result that the model 2 yields a wider density region of the pasta
phases compared with model 1
can be also understood in terms of the density-dependence parameter $L$ 
of the symmetry energy.  
Within a macroscopic model employed
in Ref.\ \cite{oyamatsuiida}
[see Eq.\ (1) and Eq.\ (4) of that reference],
the saturation density at a fixed value of $x_p$ is given
by \cite{oyamatsuiida}
\begin{eqnarray}
 \rho_{\rm s}(x_p)=\rho_0 [1-3L(1-2x_p)^2/K_0], \label{eq_rho_s}
\end{eqnarray}
where $K_0$ is the incompressibility.
This equation means smaller $L$ yields higher $\rho_{\rm s}(x_p)$ 
of asymmetric nuclear matter.
The above higher $\rho_{\rm s}(x_p=0.3)$ of model 2 than that of model 1 
shows a smaller $L$ of model 2, which is 
consistent with the results of $L$ obtained from the energy of 
pure neutron matter in Sec.\ \ref{sec:pnm_nuc}.
According to a result of a comprehensive analysis of 
Ref.\ \cite{oyamatsuiida}, there is a systematic trend that
nuclear models with smaller $L$ yields a wider density region
of the pasta phases.

\begin{center}
\begin{figure*}
\rotatebox{270}{
\resizebox{13cm}{!}
{\includegraphics{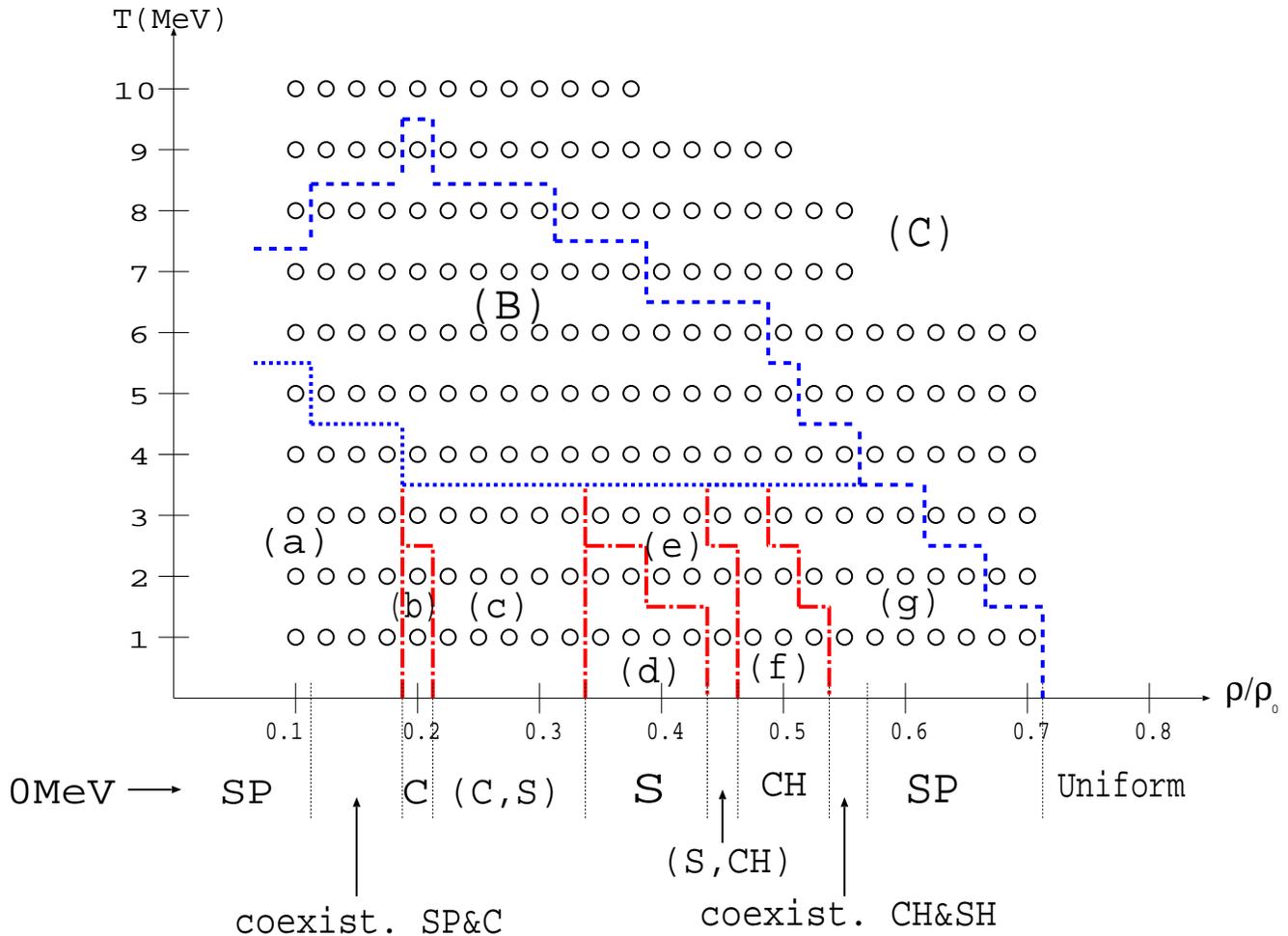}}}
\caption{ (Color online)
 Phase diagram of nuclear matter of $x_p=0.3$ at subnucear
 densities by QMD model 2 plotted in the $\rho$--$T$ plane.
 The horizontal axis is normalized in unit of the nuclear saturation
 density.
 The dashed lines correspond to phase separation lines.
 The dotted lines show
 the boundary above which nuclear surface cannot be identified.
 The dash-dotted lines show boundaries between different phases.
 Abbreviations are the same as described
 in the caption to Fig.\ \ref{phdig_cold}.
 The meanings of regions (a)--(g), (B), and (C) are explained in the
 text.
 Simulations have been carried out at the points denoted by circles.
 \label{phdig_hot}}
\end{figure*}
\end{center}

Phase diagram of model 2 for $x_p=0.3$ at nonzero temperatures
is shown in Fig.\ \ref{phdig_hot}.
Each region of this phase diagram is defined as follows: 
(a) SP, (b) C, (c) (C,S), (d) S,
(e) (S,CH), (f) CH, (g) SH, (B) phase separating region, and
(C) uniform matter.
Abbreviations SP, C, S, CH, SH, and (A,B) are the same as in
Fig.\ \ref{phdig_cold}.
Compared with the phase diagram of model 1 (see
Fig.\ 19 in Ref.\ \cite{qmd_hot}), both the pasta phases and the liquid-gas
phase separating region survive until higher temperatures:
in the case of model 1, the nuclear surface cannot be identified
above 2--3 MeV, and the critical temperature $T_{\mathrm{c}}$
of the phase separation is at $T\gtrsim 6$ MeV,
whereas for model 2, the surface melting temperature is
at 3--4 MeV and $T_{\mathrm{c}}$ is at $T\gtrsim 9$ MeV.
The increase of the surface melting temperature can be explained by
higher energy of pure neutron matter, which prevents neutrons
from dripping out of nuclei.
In addition, as described in Sec.\ \ref{sec:pnm_nuc},
the surface diffuseness for model 2 is smaller than that for model 1.
These properties keep the high density contrast between the inside and
the outside of nuclei,
and hence nuclear surface remains to be identified at higher
temperatures.
However, the increase of the critical temperature of the
phase separation for model 2 may be explained by the smaller value of
$L$, which increases the density where the proton
clustering instability takes place \cite{oyamatsuiida}.
Phase separation at high temperatures would be also induced by
the proton clustering.
Thus the same expectation may be possible for this situation,
i.e., higher symmetry energy due to the smaller value of $L$
destabilizes uniform matter against the phase separation.
As a result, the phase separation occurs at relatively higher
temperatures for model 2.

\section{Discussion and Conclusion}

We investigated the phase diagram of the pasta phases
both at zero and nonzero temperatures by QMD with two different models.
Properties of these two models are compared
by calculating the energy and the proton chemical
potential of pure neutron matter, the surface diffuseness and
the surface energy of several typical heavy nuclei.
Differences in the phase diagram, especially the expansion of the
density and temperature region of the nonuniform phases 
can be explained by these properties of pure neutron matter.
The sequence of the nuclear shape with increasing density and 
the qualitative feature of thermal fluctuation on the nucleon distribution 
with increasing temperature are the same as observed in our previous study 
\cite{qmd_hot}.
The general picture of the change of the nucleon distribution 
at a fixed density with increasing temperature is as follows.

At low temperatures, $T=1$--$1.5$ MeV for model 1 and $T=1$--$2$ MeV
for model 2; the number density of 
evaporated neutrons increases with
temperature. However, the structure of nuclei does not largely change
from that at $T=0$.
The nuclear surface becomes diffuse and
the volume fraction of nuclei increases by thermal expansion.

At intermediate temperatures, $T=1.5$--$2.5$ MeV for model 1 and
$T=2$--$3$ MeV for model 2; the nuclear shape is significantly
deformed and in some cases phase transition 
between different nuclear structures 
is triggered by the increase of the volume fraction of nuclei.
Thus the density of the phase boundary between the different nuclear shapes 
decreases with increasing temperature.

At high temperatures, $T\simeq 2.5$--$3$ MeV for model 1 and
$T\simeq3$--$4$ MeV for model 2; evaporated nucleons are dominant
and the nuclear surface can no longer be identified.
However, the long-range correlations between nucleons
due to the liquid-gas phase separation 
remain until a higher temperature
$T\simeq 6$ MeV for model 1 and $T\simeq 9$ MeV for model 2.
Above these temperatures, inhomogeneity disappears at any density.

The density-dependence parameter $L$ of the symmetry energy
is the key to understand the uncertainties of the density region
of the pasta phases in cold neutron stars \cite{oyamatsuiida}.
This parameter is also helpful to understand the present
results and to predict the general tendency of phase diagram of
the pasta phases in supernova cores.
From the energy of cold neutron matter, we obtain 
$L=93$ MeV for QMD model 1 and $L=80$ MeV for model 2.
The larger critical temperature $T_{\mathrm{c}}$
for the phase separation of model 2 would be due to the smaller value of $L$.
In addition, the smaller $L$ also yields higher saturation
density of asymmetric nuclear matter [see Eq.\ (\ref{eq_rho_s})],
which in turn increases the density
at which the system becomes uniform nuclear matter.
If one uses a nuclear force with a smaller value of $L$,
the density and temperature region of nonuniform nuclear matter
would broaden

Let us now discuss the abundance of nonuniform phases of nuclear matter 
in supernova cores.
Here we take EOS by
Shen {\it et al.} \cite{sheneos} as an example,
which is one of the widely used EOSs in supernova simulations.
For the nuclear model employed in this EOS, 
we estimate $L\simeq 120$ MeV,
which is rather high in the range of uncertainty of nuclear forces
(see, e.g., Fig.\ 1 of Ref.\ \cite{oyamatsuiida}).
In our previous work \cite{opacity}, using EOS by Shen {\it et al.}, 
we have estimated the mass fraction of the pasta phases just before bounce 
and have obtained $\sim 10$--$20 \, \%$.
Because we can expect values of $L$ for other typical EOSs 
are smaller than $L=120$ MeV for EOS by Shen {\it et al.},
our previous estimate of the mass fraction using Shen's EOS would be
very conservative.
It is reasonable to conclude that the mass fraction of
the pasta phases would be larger than $10$--$20 \, \%$
in the later stage of the collapse.

As we have shown in Ref.\ \cite{opacity},
neutrino opacity via weak neutral current in the pasta phases can be 
significantly different from that without taking account of the pasta phases.
Furthermore, even if nuclear surface melts, the neutrino opacity
of the vector current contribution
in the liquid-gas phase separating system is still larger than
that of the completely uniform gas phase.
Our present result also indicates expansion of the phase separating
region for smaller values of $L$.
Although the influence of the phase separating region on 
the mechanism of the supernova explosion
has yet to be revealed completely, it could exist not only during
collapsing phase but also after bounce.
Depending on the location of the neutrino photosphere and the
temperature profile in the late stage,
these region can affect the success of supernova explosions
\cite{com_yamada}.

\begin{acknowledgements}
We are grateful to T. Maruyama, S. Chikazumi, K. Iida, K. Oyamatsu,
and S. Yamada
for helpful discussions and comments.
This work was supported in part by JSPS,
by Grants-in-Aid for Scientific Research provided by the
Ministry of Education, Culture, Sports,
Science and Technology through research grants 14079202
and 19104006,
and by the Nishina Memorial Foundation.
Parts of the simulations were performed using MDGRAPE-2 \cite{mdgrape2}
and MDGRAPE-3 \cite{mdgrape3} of the RIKEN Super Combined Cluster
System.
\end{acknowledgements}

\bibliography{pasta_model.bbl}

\begin{thebibliography}{99}
\expandafter\ifx\csname natexlab\endcsname\relax\def\natexlab#1{#1}\fi
\expandafter\ifx\csname bibnamefont\endcsname\relax
  \def\bibnamefont#1{#1}\fi
\expandafter\ifx\csname bibfnamefont\endcsname\relax
  \def\bibfnamefont#1{#1}\fi
\expandafter\ifx\csname citenamefont\endcsname\relax
  \def\citenamefont#1{#1}\fi
\expandafter\ifx\csname url\endcsname\relax
  \def\url#1{\texttt{#1}}\fi
\expandafter\ifx\csname urlprefix\endcsname\relax\def\urlprefix{URL }\fi
\providecommand{\bibinfo}[2]{#2}
\providecommand{\eprint}[2][]{\url{#2}}

\bibitem[{\citenamefont{Ravenhall et~al.}(1983)\citenamefont{Ravenhall,
  Pethick, and Wilson}}]{rpw}
\bibinfo{author}{\bibfnamefont{D.~G.} \bibnamefont{Ravenhall}},
  \bibinfo{author}{\bibfnamefont{C.~J.} \bibnamefont{Pethick}},
  \bibnamefont{and} \bibinfo{author}{\bibfnamefont{J.~R.}
  \bibnamefont{Wilson}}, \bibinfo{journal}{Phys.\ Rev.\ Lett.}
  \textbf{\bibinfo{volume}{50}}, \bibinfo{pages}{2066} (\bibinfo{year}{1983}).

\bibitem[{\citenamefont{Hashimoto et~al.}(1984)\citenamefont{Hashimoto, Seki,
  and Yamada}}]{hashimoto}
\bibinfo{author}{\bibfnamefont{M.}~\bibnamefont{Hashimoto}},
  \bibinfo{author}{\bibfnamefont{H.}~\bibnamefont{Seki}}, \bibnamefont{and}
  \bibinfo{author}{\bibfnamefont{M.}~\bibnamefont{Yamada}},
  \bibinfo{journal}{Prog.\ Theor.\ Phys.} \textbf{\bibinfo{volume}{71}},
  \bibinfo{pages}{320} (\bibinfo{year}{1984}).

\bibitem[{\citenamefont{Jennings and Schwenk}()}]{js_review}
\bibinfo{author}{\bibfnamefont{B.~K.} \bibnamefont{Jennings}} \bibnamefont{and}
  \bibinfo{author}{\bibfnamefont{A.}~\bibnamefont{Schwenk}},
  \bibinfo{note}{nucl-th/0512013}.

\bibitem[{\citenamefont{Bethe}(1990)}]{bethe_rev}
\bibinfo{author}{\bibfnamefont{H.~A.} \bibnamefont{Bethe}},
  \bibinfo{journal}{Rev.\ Mod.\ Phys.} \textbf{\bibinfo{volume}{62}},
  \bibinfo{pages}{801} (\bibinfo{year}{1990}).

\bibitem[{\citenamefont{Burrows et~al.}(2006)\citenamefont{Burrows, Reddy, and
  Thompson}}]{burrows_review}
\bibinfo{author}{\bibfnamefont{A.}~\bibnamefont{Burrows}},
  \bibinfo{author}{\bibfnamefont{S.}~\bibnamefont{Reddy}}, \bibnamefont{and}
  \bibinfo{author}{\bibfnamefont{T.~A.} \bibnamefont{Thompson}},
  \bibinfo{journal}{Nucl.\ Phys.} \textbf{\bibinfo{volume}{A777}},
  \bibinfo{pages}{356} (\bibinfo{year}{2006}).

\bibitem[{\citenamefont{Janka et~al.}(2007)\citenamefont{Janka, Langanke,
  Merek, Mart\'{i}nez-Pinedo, and M{\"u}ller}}]{janka_review}
\bibinfo{author}{\bibfnamefont{H.-T.} \bibnamefont{Janka}},
  \bibinfo{author}{\bibfnamefont{K.}~\bibnamefont{Langanke}},
  \bibinfo{author}{\bibfnamefont{A.}~\bibnamefont{Merek}},
  \bibinfo{author}{\bibnamefont{Mart\'{i}nez-Pinedo}}, \bibnamefont{and}
  \bibinfo{author}{\bibfnamefont{B.}~\bibnamefont{M{\"u}ller}},
  \bibinfo{journal}{Phys.\ Rep.} \textbf{\bibinfo{volume}{442}},
  \bibinfo{pages}{38} (\bibinfo{year}{2007}).

\bibitem[{\citenamefont{Horowitz et~al.}(2004)\citenamefont{Horowitz,
  P\'{e}rez-Garc\'{i}a, Carriere, Berry, and Piekarewicz}}]{horowitz}
\bibinfo{author}{\bibfnamefont{C.~J.} \bibnamefont{Horowitz}},
  \bibinfo{author}{\bibfnamefont{M.~A.} \bibnamefont{P\'{e}rez-Garc\'{i}a}},
  \bibinfo{author}{\bibfnamefont{J.}~\bibnamefont{Carriere}},
  \bibinfo{author}{\bibfnamefont{D.~K.} \bibnamefont{Berry}}, \bibnamefont{and}
  \bibinfo{author}{\bibfnamefont{J.}~\bibnamefont{Piekarewicz}},
  \bibinfo{journal}{Phys.\ Rev.\ C} \textbf{\bibinfo{volume}{70}},
  \bibinfo{pages}{065806} (\bibinfo{year}{2004}).

\bibitem[{\citenamefont{Sonoda et~al.}(2007)\citenamefont{Sonoda, Watanabe,
  Sato, Takiwaki, Yasuoka, and Ebisuzaki}}]{opacity}
\bibinfo{author}{\bibfnamefont{H.}~\bibnamefont{Sonoda}},
  \bibinfo{author}{\bibfnamefont{G.}~\bibnamefont{Watanabe}},
  \bibinfo{author}{\bibfnamefont{K.}~\bibnamefont{Sato}},
  \bibinfo{author}{\bibfnamefont{T.}~\bibnamefont{Takiwaki}},
  \bibinfo{author}{\bibfnamefont{K.}~\bibnamefont{Yasuoka}}, \bibnamefont{and}
  \bibinfo{author}{\bibfnamefont{T.}~\bibnamefont{Ebisuzaki}},
  \bibinfo{journal}{Phys.\ Rev.\ C} \textbf{\bibinfo{volume}{75}},
  \bibinfo{pages}{042801(R)} (\bibinfo{year}{2007}).

\bibitem[{\citenamefont{Lorenz et~al.}(1993)\citenamefont{Lorenz, Ravenhall,
  and Pethick}}]{lrp}
\bibinfo{author}{\bibfnamefont{C.~P.} \bibnamefont{Lorenz}},
  \bibinfo{author}{\bibfnamefont{D.~G.} \bibnamefont{Ravenhall}},
  \bibnamefont{and} \bibinfo{author}{\bibfnamefont{C.~J.}
  \bibnamefont{Pethick}}, \bibinfo{journal}{Phys.\ Rev.\ Lett.}
  \textbf{\bibinfo{volume}{70}}, \bibinfo{pages}{379} (\bibinfo{year}{1993}).

\bibitem[{\citenamefont{Pethick and Thorsson}(1994)}]{peth_thor}
\bibinfo{author}{\bibfnamefont{C.~J.} \bibnamefont{Pethick}} \bibnamefont{and}
  \bibinfo{author}{\bibfnamefont{V.}~\bibnamefont{Thorsson}},
  \bibinfo{journal}{Phys.\ Rev.\ Lett.} \textbf{\bibinfo{volume}{72}},
  \bibinfo{pages}{1964} (\bibinfo{year}{1994}).

\bibitem[{\citenamefont{Kaminker et~al.}(1999)\citenamefont{Kaminker, Pethick,
  Potekhin, Thorsson, and Yakovlev}}]{kaminker}
\bibinfo{author}{\bibfnamefont{A.~D.} \bibnamefont{Kaminker}},
  \bibinfo{author}{\bibfnamefont{C.~J.} \bibnamefont{Pethick}},
  \bibinfo{author}{\bibfnamefont{A.~Y.} \bibnamefont{Potekhin}},
  \bibinfo{author}{\bibfnamefont{V.}~\bibnamefont{Thorsson}}, \bibnamefont{and}
  \bibinfo{author}{\bibfnamefont{D.~G.} \bibnamefont{Yakovlev}},
  \bibinfo{journal}{Astron.\ Astrophys.} \textbf{\bibinfo{volume}{343}},
  \bibinfo{pages}{1009} (\bibinfo{year}{1999}).

\bibitem[{\citenamefont{Gusakov et~al.}(2004)\citenamefont{Gusakov, Yakovlev,
  Haensel, and Gnedin}}]{gusakov}
\bibinfo{author}{\bibfnamefont{M.~E.} \bibnamefont{Gusakov}},
  \bibinfo{author}{\bibfnamefont{D.~G.} \bibnamefont{Yakovlev}},
  \bibinfo{author}{\bibfnamefont{P.}~\bibnamefont{Haensel}}, \bibnamefont{and}
  \bibinfo{author}{\bibfnamefont{O.~Y.} \bibnamefont{Gnedin}},
  \bibinfo{journal}{Astron.\ Astrophys.} \textbf{\bibinfo{volume}{421}},
  \bibinfo{pages}{1143} (\bibinfo{year}{2004}).

\bibitem[{\citenamefont{Williams and Koonin}(1985)}]{williams}
\bibinfo{author}{\bibfnamefont{R.~D.} \bibnamefont{Williams}} \bibnamefont{and}
  \bibinfo{author}{\bibfnamefont{S.~E.} \bibnamefont{Koonin}},
  \bibinfo{journal}{Nucl.\ Phys.} \textbf{\bibinfo{volume}{A435}},
  \bibinfo{pages}{844} (\bibinfo{year}{1985}).

\bibitem[{\citenamefont{Lassaut et~al.}(1987)\citenamefont{Lassaut, Flocard,
  Bonche, Heenen, and Suraud}}]{lassaut}
\bibinfo{author}{\bibfnamefont{M.}~\bibnamefont{Lassaut}},
  \bibinfo{author}{\bibfnamefont{H.}~\bibnamefont{Flocard}},
  \bibinfo{author}{\bibfnamefont{P.}~\bibnamefont{Bonche}},
  \bibinfo{author}{\bibfnamefont{P.~H.} \bibnamefont{Heenen}},
  \bibnamefont{and} \bibinfo{author}{\bibfnamefont{E.}~\bibnamefont{Suraud}},
  \bibinfo{journal}{Astron.\ Astrophys.} \textbf{\bibinfo{volume}{183}},
  \bibinfo{pages}{L3} (\bibinfo{year}{1987}).

\bibitem[{\citenamefont{Oyamatsu}(1993)}]{oyamatsu}
\bibinfo{author}{\bibfnamefont{K.}~\bibnamefont{Oyamatsu}},
  \bibinfo{journal}{Nucl.\ Phys.} \textbf{\bibinfo{volume}{A561}},
  \bibinfo{pages}{431} (\bibinfo{year}{1993}).

\bibitem[{\citenamefont{Watanabe et~al.}(2000)\citenamefont{Watanabe, Iida, and
  Sato}}]{watanabe_liquid_nsm}
\bibinfo{author}{\bibfnamefont{G.}~\bibnamefont{Watanabe}},
  \bibinfo{author}{\bibfnamefont{K.}~\bibnamefont{Iida}}, \bibnamefont{and}
  \bibinfo{author}{\bibfnamefont{K.}~\bibnamefont{Sato}},
  \bibinfo{journal}{Nucl.\ Phys.} \textbf{\bibinfo{volume}{A676}},
  \bibinfo{pages}{455} (\bibinfo{year}{2000}).

\bibitem[{\citenamefont{Watanabe et~al.}(2001)\citenamefont{Watanabe, Iida, and
  Sato}}]{watanabe_liquid}
\bibinfo{author}{\bibfnamefont{G.}~\bibnamefont{Watanabe}},
  \bibinfo{author}{\bibfnamefont{K.}~\bibnamefont{Iida}}, \bibnamefont{and}
  \bibinfo{author}{\bibfnamefont{K.}~\bibnamefont{Sato}},
  \bibinfo{journal}{Nucl.\ Phys.} \textbf{\bibinfo{volume}{A687}},
  \bibinfo{pages}{512} (\bibinfo{year}{2001}).

\bibitem[{\citenamefont{Watanabe
  et~al.}(2003{\natexlab{a}})\citenamefont{Watanabe, Iida, and
  Sato}}]{watanabe_liquid_err}
\bibinfo{author}{\bibfnamefont{G.}~\bibnamefont{Watanabe}},
  \bibinfo{author}{\bibfnamefont{K.}~\bibnamefont{Iida}}, \bibnamefont{and}
  \bibinfo{author}{\bibfnamefont{K.}~\bibnamefont{Sato}},
  \bibinfo{journal}{Nucl.\ Phys.} \textbf{\bibinfo{volume}{A726}},
  \bibinfo{pages}{357} (\bibinfo{year}{2003}{\natexlab{a}}).

\bibitem[{\citenamefont{Maruyama et~al.}(2005)\citenamefont{Maruyama, Tatsumi,
  Voskresensky, Tanigawa, and Chiba}}]{maru_scr}
\bibinfo{author}{\bibfnamefont{T.}~\bibnamefont{Maruyama}},
  \bibinfo{author}{\bibfnamefont{T.}~\bibnamefont{Tatsumi}},
  \bibinfo{author}{\bibfnamefont{D.~N.} \bibnamefont{Voskresensky}},
  \bibinfo{author}{\bibfnamefont{T.}~\bibnamefont{Tanigawa}}, \bibnamefont{and}
  \bibinfo{author}{\bibfnamefont{S.}~\bibnamefont{Chiba}},
  \bibinfo{journal}{Phys.\ Rev.\ C} \textbf{\bibinfo{volume}{72}},
  \bibinfo{pages}{015802} (\bibinfo{year}{2005}).

\bibitem[{\citenamefont{Cheng et~al.}(1997)\citenamefont{Cheng, Yao, and
  Dai}}]{cheng}
\bibinfo{author}{\bibfnamefont{K.~S.} \bibnamefont{Cheng}},
  \bibinfo{author}{\bibfnamefont{C.~C.} \bibnamefont{Yao}}, \bibnamefont{and}
  \bibinfo{author}{\bibfnamefont{Z.~G.} \bibnamefont{Dai}},
  \bibinfo{journal}{Phys.\ Rev.\ C} \textbf{\bibinfo{volume}{55}},
  \bibinfo{pages}{2092} (\bibinfo{year}{1997}).

\bibitem[{\citenamefont{Douchin and Haensel}(2000)}]{douchin}
\bibinfo{author}{\bibfnamefont{F.}~\bibnamefont{Douchin}} \bibnamefont{and}
  \bibinfo{author}{\bibfnamefont{P.}~\bibnamefont{Haensel}},
  \bibinfo{journal}{Phys.\ Lett.} \textbf{\bibinfo{volume}{B485}},
  \bibinfo{pages}{107} (\bibinfo{year}{2000}).

\bibitem[{\citenamefont{Watanabe et~al.}(2002)\citenamefont{Watanabe, Sato,
  Yasuoka, and Ebisuzaki}}]{qmd_rc}
\bibinfo{author}{\bibfnamefont{G.}~\bibnamefont{Watanabe}},
  \bibinfo{author}{\bibfnamefont{K.}~\bibnamefont{Sato}},
  \bibinfo{author}{\bibfnamefont{K.}~\bibnamefont{Yasuoka}}, \bibnamefont{and}
  \bibinfo{author}{\bibfnamefont{T.}~\bibnamefont{Ebisuzaki}},
  \bibinfo{journal}{Phys.\ Rev.\ C} \textbf{\bibinfo{volume}{66}},
  \bibinfo{pages}{012801(R)} (\bibinfo{year}{2002}).

\bibitem[{\citenamefont{Watanabe
  et~al.}(2003{\natexlab{b}})\citenamefont{Watanabe, Sato, Yasuoka, and
  Ebisuzaki}}]{qmd}
\bibinfo{author}{\bibfnamefont{G.}~\bibnamefont{Watanabe}},
  \bibinfo{author}{\bibfnamefont{K.}~\bibnamefont{Sato}},
  \bibinfo{author}{\bibfnamefont{K.}~\bibnamefont{Yasuoka}}, \bibnamefont{and}
  \bibinfo{author}{\bibfnamefont{T.}~\bibnamefont{Ebisuzaki}},
  \bibinfo{journal}{Phys.\ Rev.\ C} \textbf{\bibinfo{volume}{68}},
  \bibinfo{pages}{035806} (\bibinfo{year}{2003}{\natexlab{b}}).

\bibitem[{\citenamefont{Watanabe et~al.}(2004)\citenamefont{Watanabe, Sato,
  Yasuoka, and Ebisuzaki}}]{qmd_hot}
\bibinfo{author}{\bibfnamefont{G.}~\bibnamefont{Watanabe}},
  \bibinfo{author}{\bibfnamefont{K.}~\bibnamefont{Sato}},
  \bibinfo{author}{\bibfnamefont{K.}~\bibnamefont{Yasuoka}}, \bibnamefont{and}
  \bibinfo{author}{\bibfnamefont{T.}~\bibnamefont{Ebisuzaki}},
  \bibinfo{journal}{Phys.\ Rev.\ C} \textbf{\bibinfo{volume}{69}},
  \bibinfo{pages}{055805} (\bibinfo{year}{2004}).

\bibitem[{\citenamefont{Watanabe and Sonoda}(2007)}]{pasta_review}
\bibinfo{author}{\bibfnamefont{G.}~\bibnamefont{Watanabe}} \bibnamefont{and}
  \bibinfo{author}{\bibfnamefont{H.}~\bibnamefont{Sonoda}}, in
  \emph{\bibinfo{booktitle}{Soft Condensed Matter: New Research}}, edited by
  \bibinfo{editor}{\bibfnamefont{K.~I.} \bibnamefont{Dillon}}
  (\bibinfo{publisher}{Nova Science}, \bibinfo{address}{New York},
  \bibinfo{year}{2007}), p.~\bibinfo{pages}{1},
  \bibinfo{note}{cond-mat/0502515}.

\bibitem[{\citenamefont{Oyamatsu and Iida}(2007)}]{oyamatsuiida}
\bibinfo{author}{\bibfnamefont{K.}~\bibnamefont{Oyamatsu}} \bibnamefont{and}
  \bibinfo{author}{\bibfnamefont{K.}~\bibnamefont{Iida}},
  \bibinfo{journal}{Phys.\ Rev.\ C} \textbf{\bibinfo{volume}{75}},
  \bibinfo{pages}{015801} (\bibinfo{year}{2007}).

\bibitem[{\citenamefont{Watanabe et~al.}(2005)\citenamefont{Watanabe, Maruyama,
  Sato, Yasuoka, and Ebisuzaki}}]{qmd_transition}
\bibinfo{author}{\bibfnamefont{G.}~\bibnamefont{Watanabe}},
  \bibinfo{author}{\bibfnamefont{T.}~\bibnamefont{Maruyama}},
  \bibinfo{author}{\bibfnamefont{K.}~\bibnamefont{Sato}},
  \bibinfo{author}{\bibfnamefont{K.}~\bibnamefont{Yasuoka}}, \bibnamefont{and}
  \bibinfo{author}{\bibfnamefont{T.}~\bibnamefont{Ebisuzaki}},
  \bibinfo{journal}{Phys.\ Rev.\ Lett.} \textbf{\bibinfo{volume}{94}},
  \bibinfo{pages}{031101} (\bibinfo{year}{2005}).

\bibitem[{\citenamefont{Maruyama et~al.}(1998)\citenamefont{Maruyama, Niita,
  Oyamatsu, Maruyama, Chiba, and Iwamoto}}]{maruyama}
\bibinfo{author}{\bibfnamefont{T.}~\bibnamefont{Maruyama}},
  \bibinfo{author}{\bibfnamefont{K.}~\bibnamefont{Niita}},
  \bibinfo{author}{\bibfnamefont{K.}~\bibnamefont{Oyamatsu}},
  \bibinfo{author}{\bibfnamefont{T.}~\bibnamefont{Maruyama}},
  \bibinfo{author}{\bibfnamefont{S.}~\bibnamefont{Chiba}}, \bibnamefont{and}
  \bibinfo{author}{\bibfnamefont{A.}~\bibnamefont{Iwamoto}},
  \bibinfo{journal}{Phys.\ Rev.\ C} \textbf{\bibinfo{volume}{57}},
  \bibinfo{pages}{655} (\bibinfo{year}{1998}).

\bibitem[{\citenamefont{Chikazumi et~al.}(2001)\citenamefont{Chikazumi,
  Maruyama, Chiba, Niita, and Iwamoto}}]{chikazumi}
\bibinfo{author}{\bibfnamefont{S.}~\bibnamefont{Chikazumi}},
  \bibinfo{author}{\bibfnamefont{T.}~\bibnamefont{Maruyama}},
  \bibinfo{author}{\bibfnamefont{S.}~\bibnamefont{Chiba}},
  \bibinfo{author}{\bibfnamefont{K.}~\bibnamefont{Niita}}, \bibnamefont{and}
  \bibinfo{author}{\bibfnamefont{A.}~\bibnamefont{Iwamoto}},
  \bibinfo{journal}{Phys.\ Rev.\ C} \textbf{\bibinfo{volume}{63}},
  \bibinfo{pages}{024602} (\bibinfo{year}{2001}).

\bibitem[{not({\natexlab{a}})}]{note_error}
\bibinfo{note}{We remark that the expression of the Hamiltonian given in Ref.\
  \cite{chikazumi} is erroneous. The quantity $L$ ($L_\mathrm{w}^2$ in our
  notation) in the term $V_{\mathrm{Skyrme}}$ in Eq.\ (2) of Ref.\
  \cite{chikazumi} should be modified in a way as in Eq.\ (\ref{eq_exp_int}) in
  the present paper (private communication with T. Maruyama).}

\bibitem[{\citenamefont{Kido et~al.}(2000)\citenamefont{Kido, Maruyama, Niita,
  and Chiba}}]{kido}
\bibinfo{author}{\bibfnamefont{T.}~\bibnamefont{Kido}},
  \bibinfo{author}{\bibfnamefont{T.}~\bibnamefont{Maruyama}},
  \bibinfo{author}{\bibfnamefont{K.}~\bibnamefont{Niita}}, \bibnamefont{and}
  \bibinfo{author}{\bibfnamefont{S.}~\bibnamefont{Chiba}},
  \bibinfo{journal}{Nucl.\ Phys.} \textbf{\bibinfo{volume}{A663}},
  \bibinfo{pages}{877c} (\bibinfo{year}{2000}).

\bibitem[{\citenamefont{Nos\'{e}}(1983)}]{nose1}
\bibinfo{author}{\bibfnamefont{S.}~\bibnamefont{Nos\'{e}}},
  \bibinfo{journal}{J.\ Chem.\ Phys.} \textbf{\bibinfo{volume}{72}},
  \bibinfo{pages}{2384} (\bibinfo{year}{1983}).

\bibitem[{\citenamefont{Nos\'{e}}(1984)}]{nose2}
\bibinfo{author}{\bibfnamefont{S.}~\bibnamefont{Nos\'{e}}},
  \bibinfo{journal}{Mol.\ Phys.} \textbf{\bibinfo{volume}{52}},
  \bibinfo{pages}{255} (\bibinfo{year}{1984}).

\bibitem[{\citenamefont{Hoover}(1985)}]{hoover}
\bibinfo{author}{\bibfnamefont{W.~G.} \bibnamefont{Hoover}},
  \bibinfo{journal}{Phys.\ Rev.\ A} \textbf{\bibinfo{volume}{31}},
  \bibinfo{pages}{1695} (\bibinfo{year}{1985}).

\bibitem[{\citenamefont{P\'{e}rez-Garc\'{i}a}(2006)}]{nosehoover_perez}
\bibinfo{author}{\bibfnamefont{M.~A.} \bibnamefont{P\'{e}rez-Garc\'{i}a}},
  \bibinfo{journal}{J.\ Math.\ Chem.} \textbf{\bibinfo{volume}{40}},
  \bibinfo{pages}{63} (\bibinfo{year}{2006}).

\bibitem[{not({\natexlab{d}})}]{erratum}
\bibinfo{note}{There is a typographical error in Eq.\ (11) 
of Ref.\ \cite{qmd_hot}.  The factor $sp_s^2$ in that equation 
should be replaced by $s^2p_s^2$.}

\bibitem[{\citenamefont{Press et~al.}(1986)\citenamefont{Press, Teukolsky,
  Vettering, and Flannery}}]{recipe}
\bibinfo{author}{\bibfnamefont{W.~H.} \bibnamefont{Press}},
  \bibinfo{author}{\bibfnamefont{S.~A.} \bibnamefont{Teukolsky}},
  \bibinfo{author}{\bibfnamefont{W.~T.} \bibnamefont{Vettering}},
  \bibnamefont{and} \bibinfo{author}{\bibfnamefont{B.~P.}
  \bibnamefont{Flannery}}, \emph{\bibinfo{title}{Numerical Recipes}}
  (\bibinfo{publisher}{Cambridge University Press}, \bibinfo{year}{1986}).

\bibitem[{\citenamefont{Pethick et~al.}(1995)\citenamefont{Pethick, Ravenhall,
  and Lorentz}}]{nucl_mod_prl}
\bibinfo{author}{\bibfnamefont{C.~J.} \bibnamefont{Pethick}},
  \bibinfo{author}{\bibfnamefont{D.~G.} \bibnamefont{Ravenhall}},
  \bibnamefont{and} \bibinfo{author}{\bibfnamefont{C.~P.}
  \bibnamefont{Lorentz}}, \bibinfo{journal}{Nucl.\ Phys.}
  \textbf{\bibinfo{volume}{A584}}, \bibinfo{pages}{675} (\bibinfo{year}{1995}).

\bibitem[{not({\natexlab{b}})}]{note_chempot}
\bibinfo{note}{However, Fig.\ 9 of Ref.\ \cite{oyamatsuiida} shows a weak but
  systematic dependence of $\mu_p^{(0)}$ at $\rho_{n}=0.1$ fm$^{-3}$ on the
  density dependence parameter $L$ of the symmetry energy. Thus the information
  about $L$ can be extracted from the proton chemical potential. In this case,
  $\mu_p^{(0)}$ at $\rho_{n}=0.1$ fm$^{-3}$ of model 1 is slightly smaller than
  that of model 2, which suggests a smaller $L$ for model 1. This contradicts
  the conclusion obtained from the energy of pure neutron matter. However, due
  to the facts that the dependence of $\mu_p^{(0)}$ on $L$ is very weak and the
  values of $\mu_p^{(0)}$ for both the QMD models are almost the same at this
  specific density, the comparison of $L$ deduced from the energy of pure
  neutron matter is more reliable than that from $\mu_p^{(0)}$ \cite{com_iida}.
  Thus we conclude the value of $L$ for model 1 is greater than that for model
  2.}

\bibitem[{\citenamefont{Audi et~al.}(2003)\citenamefont{Audi, Wapstra, and
  Thibault}}]{nucl_emp_data}
\bibinfo{author}{\bibfnamefont{G.}~\bibnamefont{Audi}},
  \bibinfo{author}{\bibfnamefont{A.~H.} \bibnamefont{Wapstra}},
  \bibnamefont{and} \bibinfo{author}{\bibfnamefont{C.}~\bibnamefont{Thibault}},
  \bibinfo{journal}{Nucl.\ Phys.} \textbf{\bibinfo{volume}{A729}},
  \bibinfo{pages}{337} (\bibinfo{year}{2003}).

\bibitem[{\citenamefont{Nemeth and Bethe}(1968)}]{nemeth}
\bibinfo{author}{\bibfnamefont{J.}~\bibnamefont{Nemeth}} \bibnamefont{and}
  \bibinfo{author}{\bibfnamefont{H.~A.} \bibnamefont{Bethe}},
  \bibinfo{journal}{Nucl.\ Phys.} \textbf{\bibinfo{volume}{A116}},
  \bibinfo{pages}{241} (\bibinfo{year}{1968}).

\bibitem[{not({\natexlab{c}})}]{note_number}
\bibinfo{note}{In order to remove the finite boundary effects of the simulation
  box we need $\mathcal{O}(10^4)$ nucleons. However, in the previous study
  \cite{qmd_hot}, we confirmed at least the two-point correlation function of
  simulations with 2048 nucleons is not so different from that with 16384
  nucleons. Thus even in the case of 2048 nucleons the finite boundary effects
  little affect the phase diagram within the accuracy of $\Delta \rho\sim
  0.01\rho_0$ and $\Delta T \sim 1$ MeV.}

\bibitem[{\citenamefont{Baym et~al.}(1971)\citenamefont{Baym, Bethe, and
  Pethick}}]{bbp}
\bibinfo{author}{\bibfnamefont{G.}~\bibnamefont{Baym}},
  \bibinfo{author}{\bibfnamefont{H.~A.} \bibnamefont{Bethe}}, \bibnamefont{and}
  \bibinfo{author}{\bibfnamefont{C.~J.} \bibnamefont{Pethick}},
  \bibinfo{journal}{Nucl.\ Phys.} \textbf{\bibinfo{volume}{A175}},
  \bibinfo{pages}{225} (\bibinfo{year}{1971}).

\bibitem[{\citenamefont{Pethick and Ravenhall}(1995)}]{pr_review}
\bibinfo{author}{\bibfnamefont{C.~J.} \bibnamefont{Pethick}} \bibnamefont{and}
  \bibinfo{author}{\bibfnamefont{D.~G.} \bibnamefont{Ravenhall}},
  \bibinfo{journal}{Annu.\ Rev.\ Nucl.\ Part.\ Sci.}
  \textbf{\bibinfo{volume}{45}}, \bibinfo{pages}{429} (\bibinfo{year}{1995}).

\bibitem[{\citenamefont{Watanabe and Iida}(2003)}]{screening}
\bibinfo{author}{\bibfnamefont{G.}~\bibnamefont{Watanabe}} \bibnamefont{and}
  \bibinfo{author}{\bibfnamefont{K.}~\bibnamefont{Iida}},
  \bibinfo{journal}{Phys.\ Rev.\ C} \textbf{\bibinfo{volume}{68}},
  \bibinfo{pages}{045801} (\bibinfo{year}{2003}).

\bibitem[{\citenamefont{Michielsen and {De Raedt}}(2001)}]{michielsen}
\bibinfo{author}{\bibfnamefont{K.}~\bibnamefont{Michielsen}} \bibnamefont{and}
  \bibinfo{author}{\bibfnamefont{H.}~\bibnamefont{{De Raedt}}},
  \bibinfo{journal}{Phys.\ Rep.} \textbf{\bibinfo{volume}{347}},
  \bibinfo{pages}{461} (\bibinfo{year}{2001}).

\bibitem[{\citenamefont{{Gott III} et~al.}(1986)\citenamefont{{Gott III},
  Melott, and Dickinson}}]{gott}
\bibinfo{author}{\bibfnamefont{J.~R.} \bibnamefont{{Gott III}}},
  \bibinfo{author}{\bibfnamefont{A.~L.} \bibnamefont{Melott}},
  \bibnamefont{and}
  \bibinfo{author}{\bibfnamefont{M.}~\bibnamefont{Dickinson}},
  \bibinfo{journal}{Astrophys.\ J.} \textbf{\bibinfo{volume}{306}},
  \bibinfo{pages}{341} (\bibinfo{year}{1986}).

\bibitem[{\citenamefont{Weinberg}(1988)}]{weinberg}
\bibinfo{author}{\bibfnamefont{D.~H.} \bibnamefont{Weinberg}},
  \bibinfo{journal}{Publ.\ Astron.\ Soc.\ Pac.} \textbf{\bibinfo{volume}{100}},
  \bibinfo{pages}{1373} (\bibinfo{year}{1988}).

\bibitem[{\citenamefont{Ohta}(2006)}]{ohta_review}
\bibinfo{author}{\bibfnamefont{T.}~\bibnamefont{Ohta}},
  \bibinfo{journal}{Prog.\ Theor.\ Phys.\ Supp.}
  \textbf{\bibinfo{volume}{164}}, \bibinfo{pages}{203} (\bibinfo{year}{2006}).

\bibitem[{not({\natexlab{d}})}]{note_saturation}
\bibinfo{note}{These saturation densities and saturation energies at $x_p=0.3$
  are determined by calculating the energy of the ground state of nuclear
  matter ($x_p=0.3$) without the Coulomb interaction around the saturation
  density.}

\bibitem[{\citenamefont{Shen et~al.}(1998)\citenamefont{Shen, Toki, Oyamatsu,
  and Sumiyoshi}}]{sheneos}
\bibinfo{author}{\bibfnamefont{H.}~\bibnamefont{Shen}},
  \bibinfo{author}{\bibfnamefont{H.}~\bibnamefont{Toki}},
  \bibinfo{author}{\bibfnamefont{K.}~\bibnamefont{Oyamatsu}}, \bibnamefont{and}
  \bibinfo{author}{\bibfnamefont{K.}~\bibnamefont{Sumiyoshi}},
  \bibinfo{journal}{Nucl.\ Phys.} \textbf{\bibinfo{volume}{A637}},
  \bibinfo{pages}{435} (\bibinfo{year}{1998}).

\bibitem[{com({\natexlab{a}})}]{com_yamada}
\bibinfo{note}{S. Yamada, private communication} (\bibinfo{year}{2007}).

\bibitem[{\citenamefont{Susukita et~al.}(2003)\citenamefont{Susukita,
  Ebisuzaki, Elmegreen, Furusawa, Kato, Kawai, Kobayashi, Koishi, McNiven,
  Narumi et~al.}}]{mdgrape2}
\bibinfo{author}{\bibfnamefont{R.}~\bibnamefont{Susukita}},
  {\it et~al.}, \bibinfo{journal}{Comput.\ Phys.\ Commun.}
  \textbf{\bibinfo{volume}{155}}, \bibinfo{pages}{115} (\bibinfo{year}{2003}).

\bibitem[{\citenamefont{Narumi et~al.}(2006)\citenamefont{Narumi, Ohno,
  Okimoto, Koishi, Suenaga, Futatsugi, Yanai, Himeno, Fujikawa, Ikei
  et~al.}}]{mdgrape3}
\bibinfo{author}{\bibfnamefont{T.}~\bibnamefont{Narumi}},
  {\it et~al.},
  in \emph{\bibinfo{booktitle}{Proceedings of the SC06 (High Performance
  Computing, Networking, Storage and Analysis)}} (\bibinfo{address}{Tampa,
  USA}, \bibinfo{year}{2006}).

\bibitem[{com({\natexlab{b}})}]{com_iida}
\bibinfo{note}{K. Iida, private communication} (\bibinfo{year}{2007}).

\end{thebibliography}

\end{document}